# Thermal load models for the static design of steel-concrete composite girders

Ruizheng Wang [a], Kai Peng [a], Chen Peng [b], Changhao Wang [c, d]*

[a] School of highway, Chang'an University, Xi'an 710054, China

[b] School of Civil Engineering, Tongji University, Shanghai 200092, China

[c] Department of Chemistry, College of Chemistry and Chemical Engineering, Xiamen University, and Discipline of Intelligent Instrument and Equipment, Xiamen 361005, China

[d] School of Materials Science & Engineering, Zhejiang University, Hangzhou 310027, China

∗ Corresponding author

E-mail address: wwcivilT@outlook.com (C. Wang).

**Abstract**

Although the recommended temperature gradient models of composite girders are considered in current specifications for classifying temperature effects in various countries, they are not appropriate enough for static design. Moreover, existing national standards cannot explain the mechanism of the thermal effect on the bridge. To further investigate thermal load models of composite girders, this work proposed a decomposing method for vertical nonlinear temperature gradients based on thermal effects and a calculating method for the thermal stress of composite girders. Equivalent temperature (equivalent uniform temperature, equivalent linear temperature, and equivalent nonlinear temperature), temperature difference, and cyclic equivalent uniform temperature are analyzed to reflect the characteristics of thermal effect in composite girders. The stand values of temperature differences and equivalent temperature with a 50-year return period were investigated via probabilistic statistical analysis. Additionally, two vertical thermal load models (VTLM I and VTLM II) were set up to facilitate the design and applied to stress analysis. The result demonstrates that the proposed thermal load model is more suitable than the Chinese Specification for calculating the thermal effects of composite girders.

# 0 Introduction

Bridges are subjected to periodic thermal loads that can lead to thermal self-stress, secondary thermal stress, and thermal deformation, similar to static and dynamic loads, the above three phenomena can damage the bridge structures [1–3]. A large temperature difference between the concrete deck and steel girder along the steel girder depth can cause interface shear force, relative slippage, and large thermal stress [4, 5]. Thermal stress is one of the main factors causing the crack of concrete bridge decks [6, 7], and the thermal loads of composite girders can sometimes cause thermal stress equivalent to live load [8]. Therefore, a comprehensive understanding of temperature effect mechanism and structural behavior in composite girders is greatly important to new types/designs and construction of composite girders and the evaluation of the bridge's lifetime.

Most studies focused on the extreme temperature distribution and thermal stress of bridges in different regions in the past 20 years [9–12]. The numerical analysis models of the temperature field distribution of the bridge were established and analyzed by Mirambell and Aguado [13], which indicated that spatial temperature can be simplified to a 1D temperature distribution. Roberts-Wollman et al. [14] studied the thermal stress of the concrete box girder by conducting a field experiment and the result demonstrated that the thermal stress depends largely on the form of temperature distribution. Zuk et al. [15] carried out a field test on a steel-concrete composite girder to investigate temperature gradient and found that the extreme temperature difference is 22°C. Wang et al. [16] built a temperature gradient model for steel box girders based on long-term monitoring data. To further simplify the calculation of thermal stress, many researchers proposed various simplified temperature difference gradient patterns for the design of bridge structures, such as multiple parabolic forms [17] and multisegment line forms [18–20]. Thermal loads are classified into uniform temperature and temperature difference gradients [21–24]. However, the temperature difference gradient models of steel-concrete composite girder bridges in JTG/T D65-06—2015 are mainly modified based on the temperature difference gradient of AASHTO, which considered the characteristics of China's climate. Additionally, the temperature gradient models of AASHTO were obtained from the statistical analysis of thermal loads in concrete bridges. Therefore, the applicability of the temperature difference gradient modes for composite girder bridges in China must be further discussed.

Existing national standards divide the thermal effect into uniform temperature and temperature gradient according to environmental factors [25–29]. Although this classification is easy to understand, it can not explain the mechanism of the thermal effect on the bridge. Uniform temperature causes axial expansion deformation in the statically determinate bridge structure and axial secondary stress in the statically indeterminate bridge structure [30]. In the statically determinate structure, the temperature gradient can cause a small part of axial expansion deformation, bending deformation, and self-balanced thermal self-stress [31, 32]. In the statically indeterminate structure, the temperature gradient can generate axial secondary stress, thermal self-stress, and bending secondary stress [33, 34]. The prior temperature gradient modes are mixed with the secondary thermal effect and self-generated thermal effect. Therefore, these temperature gradient frameworks ignore the mechanical principle of the temperature effect.

To solve the above issues, this work aimed to investigate thermal load models for designing steel-concrete composite girders and proposed an innovative calculation method for decomposing vertical nonlinear thermal loads based on the thermal effect equivalence principle. A systematic approach was established for two vertical thermal loads, and long-term data of temperature field distribution in composite girder were collected and analyzed. Temperature difference and equivalent temperature with a 50-year recurrence period were determined using the probability limit-state method. Finally, two vertical thermal load models (VTLMs) were established, and a calculation method for thermal stress was proposed. Thermal stress was compared between the designed thermal load model and the recommendations in Chinese Specification (CS), and the relationship between the thermal effect and loads was also examined.

## 1 Innovative method to establish a design thermal load model

### 1.1 Decomposition of vertical nonlinear temperature gradient

Suppose the vertical nonlinear temperature gradient curve of the steel-concrete composite girder cross-section is $T(y)$ and the origin of the vertical coordinate ($y$-axis) is taken at the centroid of the converted section of steel-concrete composite girder with a positive value toward an upward direction, the free deformation of the section with thermal variation is expressed as $\alpha(y)T(y)$ (taking per unit length of the girder for analysis) as shown in Fig. 1. The stress caused by the free deformation of the section can be expressed as follows:

$$\sigma_T(y) = E(y)\alpha(y)T(y) \tag{1}$$

where $E(y)$ is Young's modulus of the materials at $y$ position and $\alpha(y)$ is the thermal expansion coefficient of the materials at $y$ position.

Considering self-stress and secondary stress caused by thermal loads, the overall thermal stress of the composite girder can be determined as follows:

$$\sigma_{sum}(y) = E(y)\varepsilon_{sum}(y) \tag{2}$$

where $\varepsilon_{sum}(y)$ is the total strain of the selected section containing the sectional fiber constraints and structural boundary constraints.

The overall stress at any position also can be presented as follows based on the plane section assumption:

$$\sigma(y) = E(y)(\varepsilon_0 + \varphi y) \tag{3}$$

where $\varepsilon_0$ is the axial strain at the centroid of the cross-section considering the sectional fiber constraints and structural boundary constraints. $\varphi$ and $y$ denote the curvature of the whole section and the depth of object fiber to the centroid of the converted section, respectively.

Additionally, the overall stress caused by thermal loads at height $y$ is as follows:

$$\sigma(y) = \sigma_T(y) + \sigma_{sum}(y) \tag{4}$$

Simultaneously solving equations (3) and (4) reveals the relationship between the total stress $\sigma_{sum}(y)$ and the vertical temperature gradient curve $T(y)$ as follows:

$$\sigma_{sum}(y) = E(y)[\varepsilon_0 + \varphi y - \alpha(y)T(y)] \tag{5}$$

According to the relationship between the stress and the internal force, equations (6) and (7) can be obtained.

$$N = \int_{h_1}^{h_2} \sigma_{sum}(y) b(y) dy \tag{6}$$

$$M = \int_{h_1}^{h_2} \sigma_{sum}(y) y b(y) dy \tag{7}$$

where $b(y)$ is the width at height $y$, $h_1$ is the distance from the origin to the top surface of the concrete deck, and $h_2$ is the distance from the origin to the bottom of the steel girder floor.

Here, the bending moment and axial force caused by the free deformation of the section under arbitrary temperature distribution are defined as the nominal bending moment and the nominal axial force, respectively, which can be expressed as follows:

$$N_T = \int_A \sigma_T dA = \int_{h_1}^{h_2} E(y)\alpha(y)T(y)b(y)dy \tag{8}$$

$$M_T = \int_A y\sigma_T dA = \int_{h_1}^{h_2} E(y)\alpha(y)T(y)b(y)y dy \tag{9}$$

The thermal expansion coefficient of the transformed section is as follows:

$$\alpha_0 = \frac{(\alpha_s + \alpha_c)}{2} \tag{10}$$

As shown in Fig. 1, the relationship between nominal axial force and the equivalent uniform temperature of the cross-section with arbitrary nonlinear temperature gradient can be expressed as follows:

$$N_T = E_s \alpha_0 T_T A_0 \tag{11}$$

where $T_T$ is the equivalent uniform temperature of the composite girder. $A_0$ is the area of the conversion section denoted as $A_0 = A_c / \alpha_{Es} + A_s$, where $A_c$ represents to the sectional area of the concrete deck; $A_s$ is the sectional area of the steel girder; and $\alpha_{Es} = E_s / E_c$, where $E_s$ is Young's modulus of the steel and $E_c$ is Young's modulus of the concrete.

The curvature of the conversion section caused by the nominal bending moment with a nonlinear temperature gradient is equal to that of the composite girder caused by the equivalent linear temperature gradient. $\varphi$ can be obtained as follows:

$$\varphi = \frac{\alpha_0 T_L}{H} = \frac{M_T}{E_s I_0} \tag{12}$$

where $T_L$ is the equivalent linear temperature, $I_0$ is the inertial moment of the transformed section, and $H$ is the height of the steel-concrete composite girder.

After substituting equation (8) into equation (11), $T_T$ can be obtained:

$$T_T = \frac{\int_{h_1}^{h_2} E(y)T(y)b(y)dy}{E_s A_0} \tag{13}$$

After substituting equation (9) into equation (12), $T_L$ can be obtained:

$$T_L = \frac{H\int_{h_1}^{h_2} E(y)T(y)yb(y)dy}{E_s I_0} \tag{14}$$

The equivalent linear temperature gradient curve is a linear function of special slope $k$ (Fig. 1), which is named as induced thermal linear slope and can be calculated by the equation (15):

$$k = \frac{\int_{h_1}^{h_2} E(y)T(y)yb(y)dy}{E_s I_0} \tag{15}$$

Additionally, the equivalent linear temperature gradient is denoted as $G(y)$, which can be calculated byhe following equation:

$$G(y) = k(y_0 - y) \tag{16}$$

where $y_0$ is the distance between the section neutral axis and the top surface of the bridge deck.

$\delta u$, $\delta w$, and $\delta v$ can be expressed as follows (Fig. 1):

$$\begin{aligned} \delta u &= \alpha(y)T_T \\ \delta w &= \alpha(y)ky \\ \delta v &= \alpha(y)T(y) \end{aligned} \tag{17}$$

Furthermore, the primary thermal strain of the composite girder can be calculated as follows:

$$\delta l = \alpha(y)T_{NL}(y) \tag{18}$$

where $T_{NL}(y)$ is the equivalent nonlinear temperature.

Given that the bridge is in the linear elastic phase under thermal loads, equation (19) can be obtained according to the superposition principle:

$$\delta v = \delta u + \delta w + \delta l \tag{19}$$

The equivalent nonlinear temperature gradient curve can be written as follows:

$$T_{NL}(y) = T(y) - T_T - ky \tag{20}$$

The authentic temperature gradient of the composite girder can be expressed as follows:

$$T(y) = T_T + G(y) + T_{NL}(y) \tag{21}$$

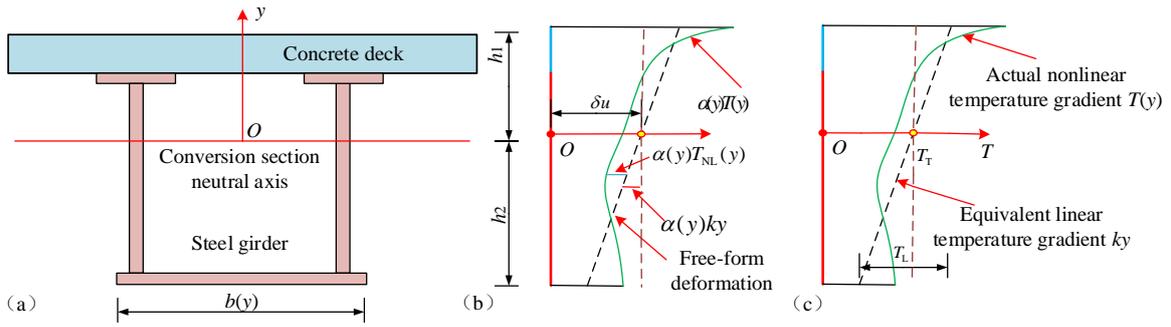

Fig. 1 Temperature distribution and thermal strain (a) section of a composite girder, (b) thermal-induced strain, and (c) temperature distribution.

**1.2 Analysis process**

After the temperature distribution of the composite girder is obtained throughout the monitoring period, $T_T$, $T_{NL}(y)$, and $T_L$ can be calculated based on the theory in Section 1.1. The temperature difference gradient can also be determined by fitting the monitored vertical temperature difference curve. The probability density function (PDF) of $T_T$, $T_L$, and the temperature difference can be statistically analyzed. Accordingly, the stand value of $T_T$, $T_{NL}(y)$, $T_L$, and the temperature difference can be proposed based on probabilistic limit-state design. Next, the temperature difference gradient of cross-section is used as VTLM I, and a combination of $T_T$, $T_L$, and $T_{NL}(y)$ is adopted as VTLM II. For further exploration of the relationship between thermal loads and thermal effects, thermal stress is analyzed based on VTLM I, VTLM II, and collected temperature data. The analysis flow of this article is shown in Fig. 2.

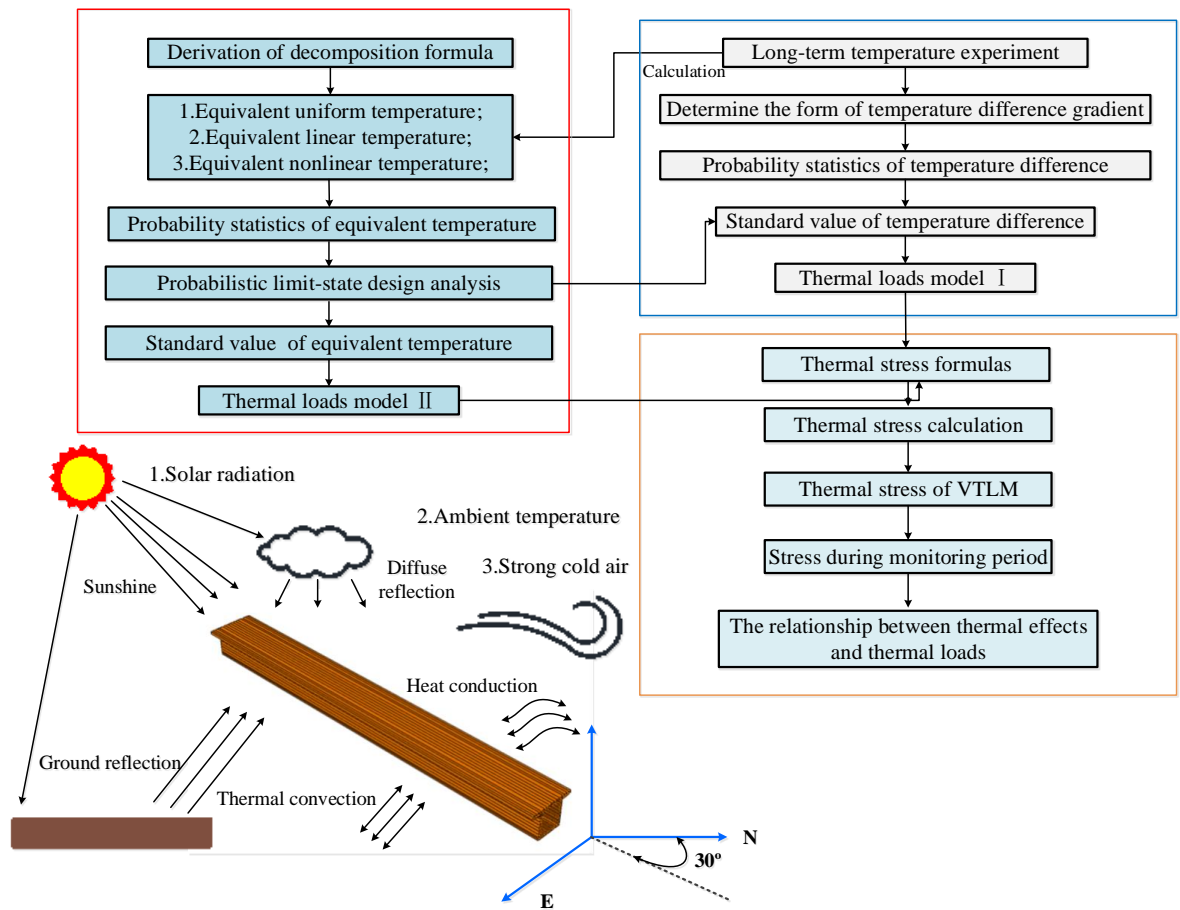

Fig. 2 Innovative method for thermal load model.

## 2 Temperature field tests

### 2.1 Sectional model and measuring point arrangement

The experimental site is located at Lanzhou New Area, Gansu Province, China, and is the intersection of the Tibetan Plateau, Mongolian Plateau, and Loess Plateau with an average altitude of 2000 m. The annual average temperature is 6.9°C. The highest temperature is in July with a monthly average of 22.5°C, and the lowest temperature is in January with a monthly average of −5.6°C. Precipitation is mainly concentrated from May to October.

The sectional model of the composite girder is 3.0 m in length and 1.45 m in depth (center line). The width of the concrete deck and bottom plate of the steel girder is 6.25 and 3.30 m. the height of the bridge deck is 0.25 m. The layouts of the measuring points at the cross-section are shown in Fig. 3 and Table 1.

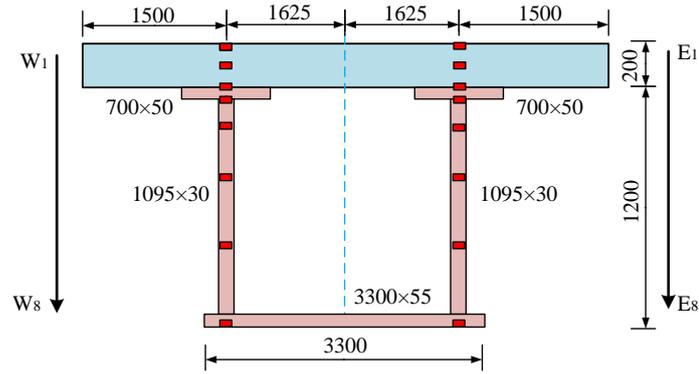

Fig. 3 Arrangement of measuring points.

Table 1 Measure locations.

| Sensor number | | Distance from the top of the bridge deck (m) |
| --- | --- | --- |
| $W_1$ | $E_1$ | 0.0 |
| $W_2$ | $E_2$ | 0.12 |
| $W_3$ | $E_3$ | 0.20 |
| $W_4$ | $E_4$ | 0.25 |
| $W_5$ | $E_5$ | 0.35 |
| $W_6$ | $E_6$ | 0.50 |
| $W_7$ | $E_7$ | 0.95 |
| $W_8$ | $E_8$ | 1.40 |

**2.2 Temperature characteristics of composite girder**

Fig. 4 exhibits the daily extreme temperature of the bridge deck and steel girder throughout the test period. It is can be seen that the daily extreme temperature shows periodic cycle from the Fig. 4. The daily maximum temperature does not exceed 50°C, and the extreme value of the daily temperature is not less than −10°C. The daily maximum temperature of the floor sometimes exceeds the daily maximum temperature of the roof. However, the daily extreme temperature can be classified as nonstationary time series.

The diurnal temperature amplitude is large because of the alternation of day and night and the change in weather conditions. Figs. 5 (a) and (b) denote the 24 h temperature cloud chart of the composite girder cross-section on July 15, 2021. These images show the temperature distribution along the roof, web, and floor of the cross-section and the change of the vertical temperature gradient with time. The influence of the solar radiation and ambient temperature on the composite girder is quite severe. The high-temperature distribution areas are mainly concentrated on the concrete deck and the bottom plate of the steel girder. The vertical temperature difference of the west web is larger than that of the east web, and the overall temperature of the west web is higher than that of the east web. The maximum positive temperature difference appears at 13:00, and the maximum negative temperature difference appears at 5:00. The statistical results also illustrate this phenomenon.

Fig. 6 (a) shows the calculation results of $T_T$ and $T_L$. $T_T$ and $T_L$ possess the typical characteristic of periodic change, and the change in $T_L$ is more stable than that in $T_T$. $T_L$ has positive and negative values, indicating that the equivalent linear temperature gradient curve may be a negative or a positive function.

The calculation results of $T_{NL}$ are shown in Fig. 6 (b). The maximum and minimum $T_{NL}$ of $E_1$ measured point are 7.9°C and −10.8°C, respectively, and those of $E_2$ measured point are 1.9°C and −2.3°C, respectively. The absolute value of $T_{NL}$ on the top concrete slab is higher than that at the bottom steel girder, so the temperature difference in $T_{NL}$ is evident.

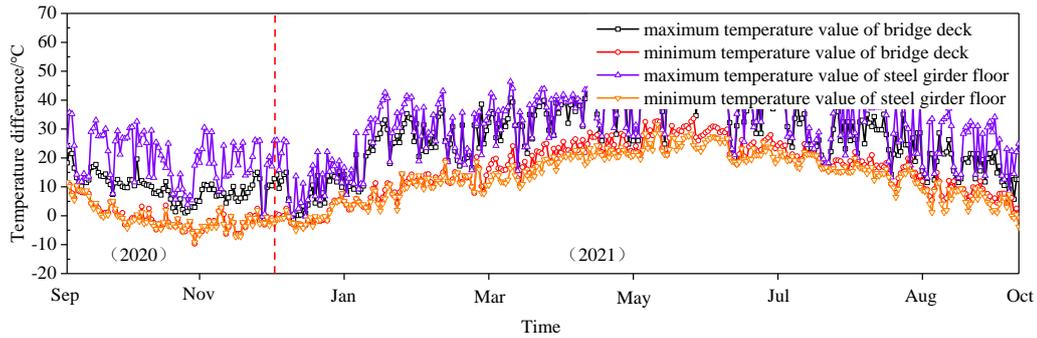

Fig. 4 Daily extreme temperature of the composite girder.

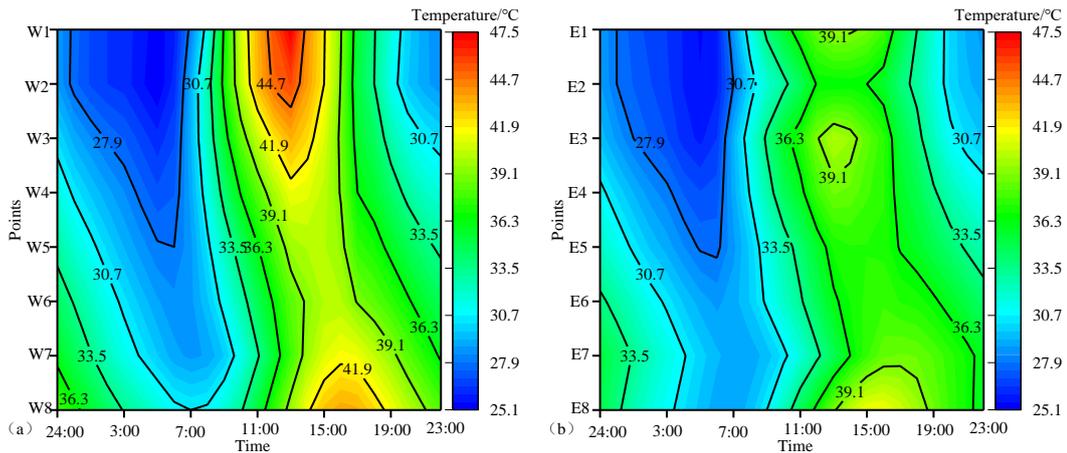

Fig. 5 A 24 h temperature cloud chart on July 15, 2021: (a) west side test point and (b) east side test point.

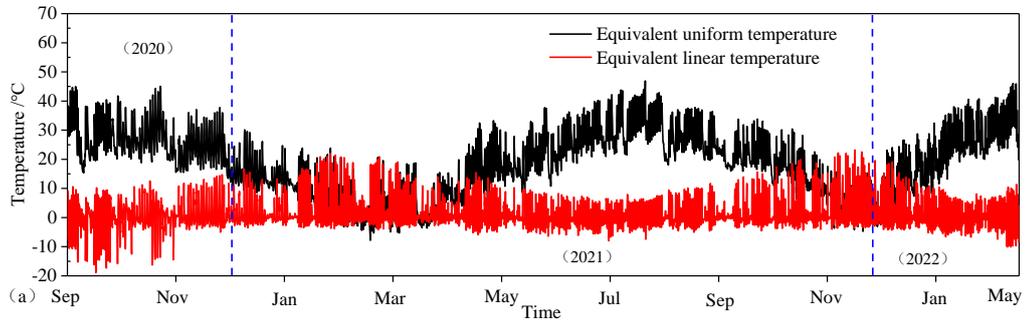

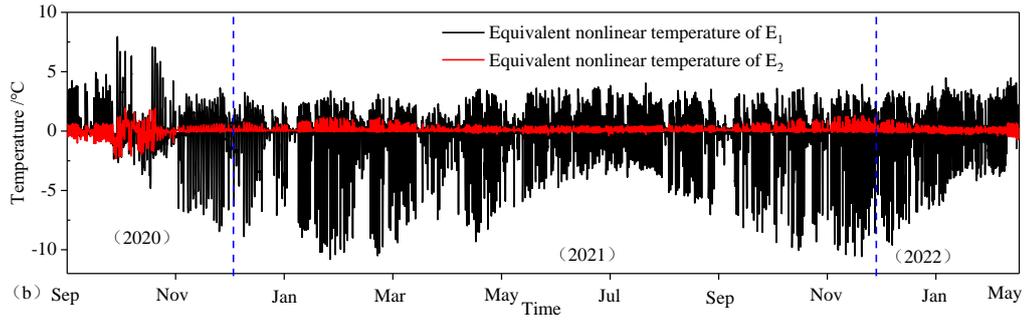

Fig. 6 (a) $T_T$, $T_L$, and (b) $T_{NL}$.

## 2.3 Temperature distribution analysis

### 2.3.1 Measured vertical thermal load model

Figs. 7 (a–f) show the measured vertical temperature difference distribution, equivalent temperature gradients, and corresponding simplified profiles. The positive and negative temperature difference gradients can be fitted by a piecewise linear function (Figs. 7(a–b)), but the function forms of equivalent positive and negative nonlinear temperature gradients are complicated (Figs. 7(c–d)). The positive and negative temperature difference gradients can be simplified as a mathematical equation. The equivalent positive and negative linear temperature gradients jointly show linear variation (Figs. 7(c–d)). However, the linear functional relationship is still unclear (Fig. 7f). On October 2, 2021, the equivalent linear temperature gradient and induced thermal scope are opposite to the sign of the corresponding temperature gradients. This phenomenon is common for bridges installed with a short web height ratio between the composite girder and the large cantilever. However, theories of probability and statistics must be used to obtain the definite expression of the equivalent temperature gradient.

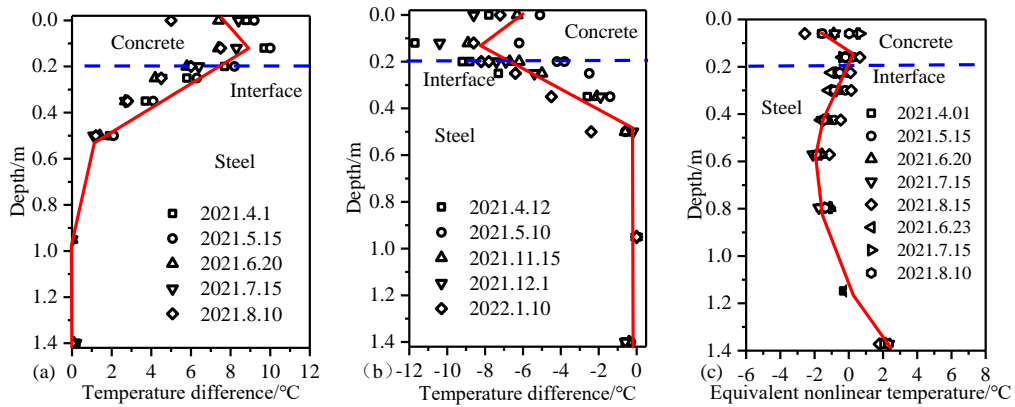

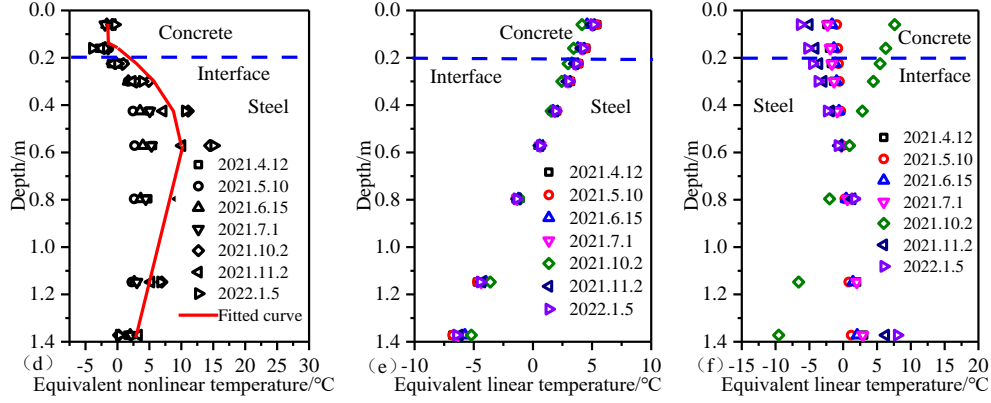

Fig. 7. Vertical thermal load model recorded in experiments: (a) positive temperature difference gradient, (b) negative temperature difference gradient, (c) equivalent positive nonlinear temperature gradient, (d) equivalent negative nonlinear temperature gradient, (e) equivalent positive linear temperature gradient, and (f) equivalent negative linear temperature gradient.

**2.3.2 Profiles of vertical temperature difference gradient**

Thermal loads of the composite girders can be divided into uniform temperature and temperature difference gradient. The uniform temperature will lead to the axial expansion deformation of the bridge, which is an essential factor in selecting appropriate expansion joints, bearings, and piers. Additionally, the uniform temperature of the composite girder can be obtained by equation (12). The detailed analysis process is shown in Section 3.3.1.

In this investigation, two VTLMs (vertical thermal load models) are proposed (Fig. 8, 15) (Figs. 8 (a–b)). According to the measured data, the profiles 1 (PVTG) and 2 (NVTG) for the thermal load model I can be obtained by the following equations:

Profile 1 (positive vertical temperature gradient):

$$T_{profile1}(y) = \begin{cases} \dfrac{2(T_{P2}-T_{P1})}{h_c}y + T_{P1}, & (0 \leq y \leq \dfrac{h_c}{2}) \\ \dfrac{T_{P2}-T_{P3}}{0.5-h_c/2}y + \dfrac{T_{P2}-(2T_{P2}-T_{P3})h_c}{1-h_c}, & (\dfrac{h_c}{2} < y \leq 0.5\text{m}) \\ \dfrac{T_{P3}}{2}(1-y), & (0.5\text{m} < y \leq 1.0\text{m}) \\ 0, & (1.0\text{m} < y \leq h) \end{cases} \quad (22)$$

where $y$ is the distance from the top surface of the bridge deck to the calculated point, m. $T_{p1}$–$T_{p3}$ are the base numbers of positive temperature differences, and $h_c$ is the height of the concrete deck.

Profile 2 (Negative vertical temperature gradient):

$$T_{profile2}(y) = \begin{cases} \dfrac{2(T_{N2}-T_{N1})}{h_c}y + T_{N1}, & (0 \leq y \leq \dfrac{h_c}{2}) \\ 0, & (0.5\text{m} < y \leq h) \\ \dfrac{T_{N2}}{0.5-h_c/2}(0.5-y), & (\dfrac{h_c}{2} < y \leq 0.5\text{m}) \end{cases} \quad (23)$$

where $T_{N1}$ and $T_{N2}$ are the base numbers of negative temperature difference.

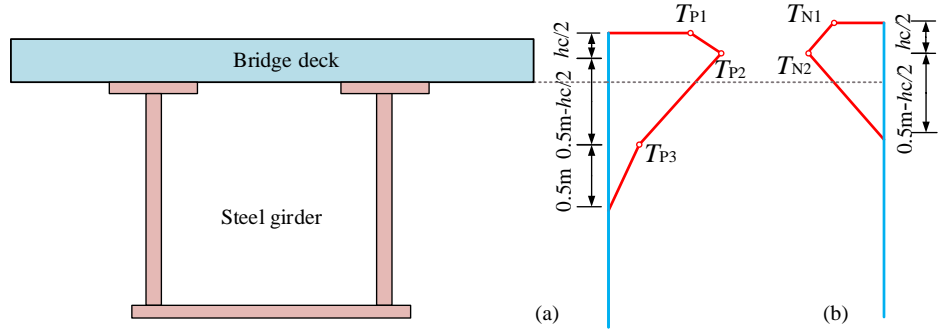

Fig. 8. Vertical thermal load model I: (a) Profile 1 and (b) Profile 2.

Thermal load model I can be regarded as the temperature difference distribution that represents the relative value of the temperature load. The VTLM I can be obtained through the statistical and reliability analyses of the temperature difference. The temperature gradient profiles composed of equivalent uniform temperature, equivalent nonlinear temperature gradient, and equivalent linear temperature gradient are named as VTLM II. VTLM II is obtained from the probabilistic analysis of the equivalent temperature (equivalent uniform temperature, equivalent linear temperature, and equivalent nonlinear temperature) of the composite girder, which is different from the analysis method for thermal load model I. However, the thermal effects of the two thermal load models are guaranteed to be consistent. Thermal load model I is reasonable because the temperature effect caused by thermal load model II is consistent with that caused by thermal load model I at any point.

The calculated daily extreme values of $T_{P1}$–$T_{P3}$ and $T_{N1}$–$T_{N2}$ are shown in Fig. 9. The daily extreme value of each temperature difference exhibits a different annual distribution. These seasonal distribution deviations are mainly due to the influence of ambient temperature and solar radiation.

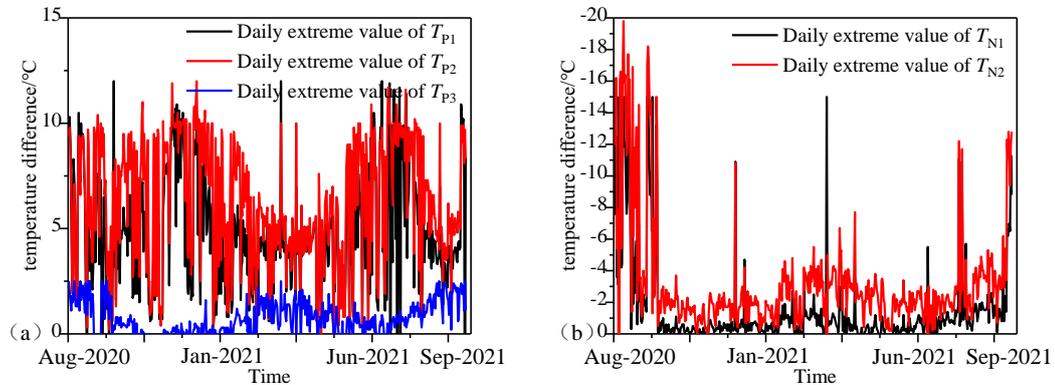

Fig. 9. Daily extreme values: (a) daily extreme values of $T_{P1}$–$T_{P3}$ and (b) daily extrmevalues of $T_{N1}$–$T_{N2}$.

## 3 Thermal load model

### 3.1 Cycling variation of equivalent uniform temperature

In the design of a long-span steel-concrete composite girder, the maximum and minimum equivalent uniform temperatures can be used to predict the degree of expansion and contraction in composite girders. a cyclic equivalent uniform temperature in the thermal load model is proposed to quantify the cumulative displacement of the long-span composite girders.

For the measured data of August 5, 2021 (Fig. 10 (a)), four key equivalent uniform temperatures are identified, namely, first temperature $t_1$, lowest temperature $t_2$, highest temperature $t_3$, and last temperature $t_4$. The daily cycling variation of the day is composed of one heating process and two cooling processes (Fig. 10 (a)) with values of 11.7°C, −6.1°C, and −6.4°C, respectively. The daily cycling variation of August 5, 2021, is 24.2°C. Fig. 10 (b) shows the frequency chart of the cyclic equivalent uniform temperature during a 300 day monitoring period, and the total daily cycling variation value is 6885°C. The cyclic equivalent uniform temperature $T_{CUT}$ of one-year can be obtained as follows:

$$T_{CUT} = \frac{365}{300} \times 6885 = 8376 \,°C \tag{24}$$

The cyclic equivalent uniform temperature $T_{CUT}$ can be considered with live loads to predict total axial displacement of bearings. Using the results as the basis, their conditions during the service period can be evaluated.

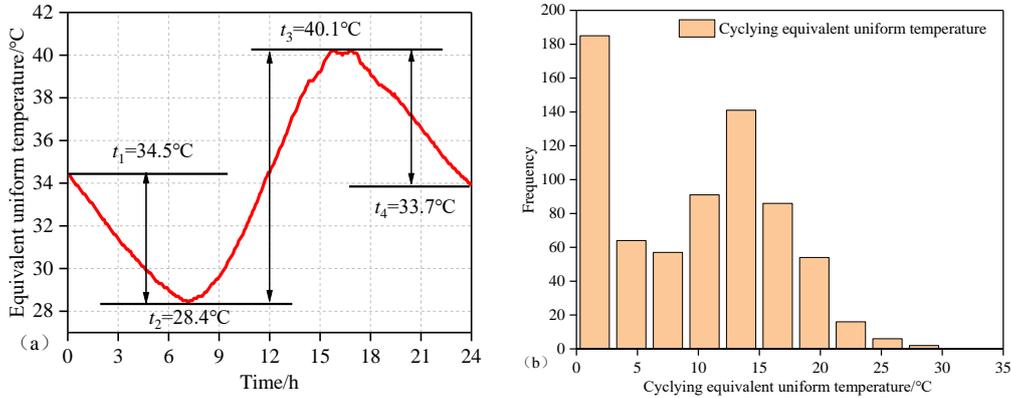

Fig. 10. (a) History of equivalent uniform temperature on August 5, 2021, and (b) frequency chart of the cyclic equivalent uniform temperature.

### 3.2 Vertical thermal load model I

After comparing the goodness of fit of multiple distribution models, the normal distribution function is finally used to describe the probability distribution of temperature difference [35, 36]. The PDF of the temperature difference distribution can be expressed as follows [37]:

$$h(TD, \mu, \sigma) = \frac{1}{\sqrt{2\pi}\sigma} \exp\left[-\frac{(TD-\mu)^2}{2\sigma^2}\right] \tag{25}$$

where $TD$ is the temperature difference of the composite girder.

The extreme values of temperature difference $TD_e$ of composite girder can be obtained by

$$P = 1 - H(TD_e, \mu, \sigma) = \int_{TD_e}^{+\infty} h(TD, \mu, \sigma) dTD \tag{26}$$

where $P$ is exceeding probability. The exceeding probability of a designed reference period within 100 years is 2%.

The frequency distribution histogram and simulated PDFs of $T_{P1}$ and $T_{N1}$ are chosen as a representative and plotted in Figs. 11(a–b). $T_{P1}$ and $T_{N1}$ fit the normal distribution. According to the probability analysis, the probability density curves can be fitted for $T_{P1}$–$T_{P3}$ and $T_{N1}$–$T_{N2}$. The parameters of fitted normal distribution and the extreme values of $T_{P1}$–$T_{P3}$ and $T_{N1}$–$T_{N2}$ are plotted as shown in Figs. 12 (a–b) and Table 2. The standard value of temperature difference of $T_{P1}$–$T_{P3}$ and $T_{N1}$–$T_{N2}$ are 16.8°C, 13.5°C, 4.2°C, −14.5°C, and −20.3°C.

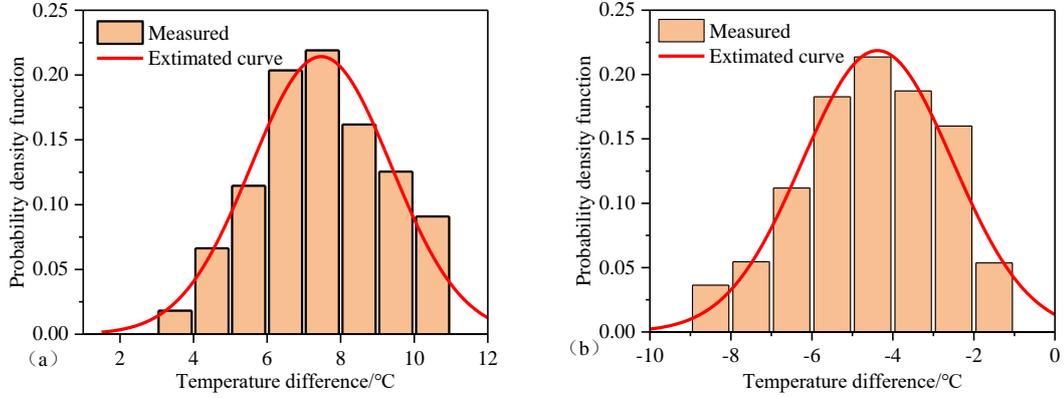

Fig. 11. Frequency distribution histogram and simulated PDFs: (a) $T_{P1}$ and (b) $T_{N1}$.

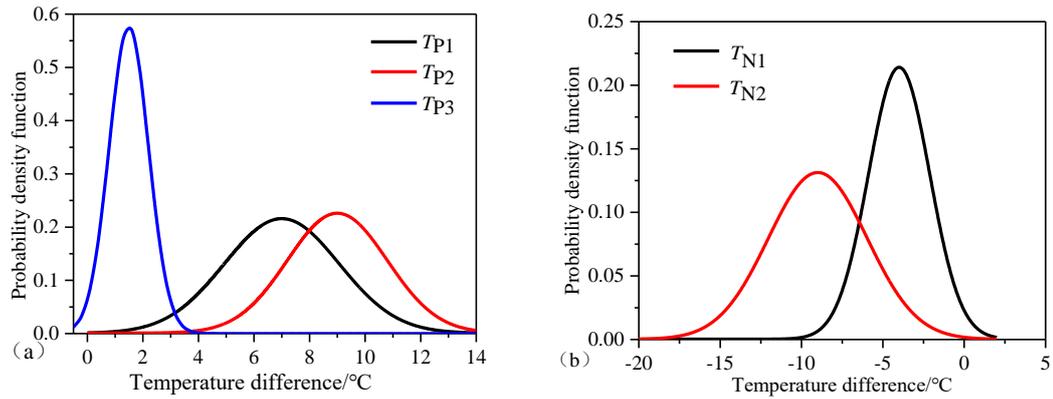

Fig. 12. Fitted PDFs: (a) $T_{P1}$–$T_{P3}$ and (b) $T_{N1}$–$T_{N2}$.

Table 2. Parameters of normal distribution and standard values within the 50-year return period.

| VTLM I | Temperature difference | Parameters | | Standard value/°C |
|---|---|---|---|---|
| | | $\mu$ | $\sigma$ | |
| Profile 1 | $T_{P1}$ | 7.471 | 1.905 | 16.8 |
| | $T_{P2}$ | 8.566 | 1.764 | 13.5 |
| | $T_{P3}$ | 1.497 | 0.712 | 4.2 |
| Profile 2 | $T_{N1}$ | −4.380 | 1.857 | −14.5 |
| | $T_{N2}$ | −9.103 | 3.043 | −20.3 |

## 3.3 Vertical thermal load model II

### 3.3.1 Equivalent uniform temperature and equivalent linear temperature

Equivalent uniform temperature provides essential information for steel-concrete composite designs following the specifications. Fig. 5 indicates that the temperature distribution on the composite girder section is non-uniform. Therefore, rather than individual temperature data, the equivalent uniform temperature of the whole section should be used to predict the thermal effect. Equivalent linear temperature gradient is the main factor that causes the secondary effect of statically indeterminate structure. The most unfavorable values (the stand values) of $T_T$ and $T_L$ in the design reference period are determined by probability statistics based on the requirements of the probabilistic limit-state design.

Gaussian mixture model (GMM) is trained by an expectation maximization algorithm. In theory, it can fit any type of distribution and is usually used to solve the problem, that is, the data under the same set contain multiple different distributions [38, 39]. The PDF of GMM can be expressed as follows:

$$f(T) = \sum_{i=1}^{M} \omega_i \frac{1}{\sqrt{2\pi}\sigma} \exp[\frac{-(T-\mu_i)^2}{2\sigma_i^2}] \qquad (27)$$

where $T$ is the equivalent temperature of the calculated point; $\mu_i$ and $\sigma_i$ are parameters to be estimated; $\omega_i$ and $M$ are the weight and number of Gaussian component, respectively. Additionally, $\sum_{i=1}^{M} \omega_i = 1$.

Equation (27) is transformed into equation (28) for operation, and the least square method is used to estimate the unknown parameters $M$, $\mu_i$, $a_i$, and $b_i$.

$$f(T) = \sum_{i=1}^{M} a_i \exp[\frac{-(T-\mu_i)^2}{b_i^2}] \qquad (28)$$

Fitting goodness is compared among multiple PDFs. The best fitting goodness is obtained when the component number ($M$) is 5, 2, 3, and 2 for $T_T^+$, $T_L^+$, $T_T^-$, and $T_L^-$, respectively. Figs. 13 (a–d) show the probability density histograms of negative and positive equivalent uniform temperature, negative and positive equivalent uniform temperature, and the estimated curve. The estimated probability density curves can accurately reflect the statistical characteristics of the probability density of the calculated temperature samples. Table 3 lists the estimated parameters of the GMM for the equivalent temperature at each calculated point during the monitoring period.

The standard equivalent linear temperature with a 50-year return period is 27.5°C and −12.3°C caused by the positive and negative temperature gradients, respectively, with the corresponding equivalent uniform temperature of 52.4°C and −13.5°C, respectively.

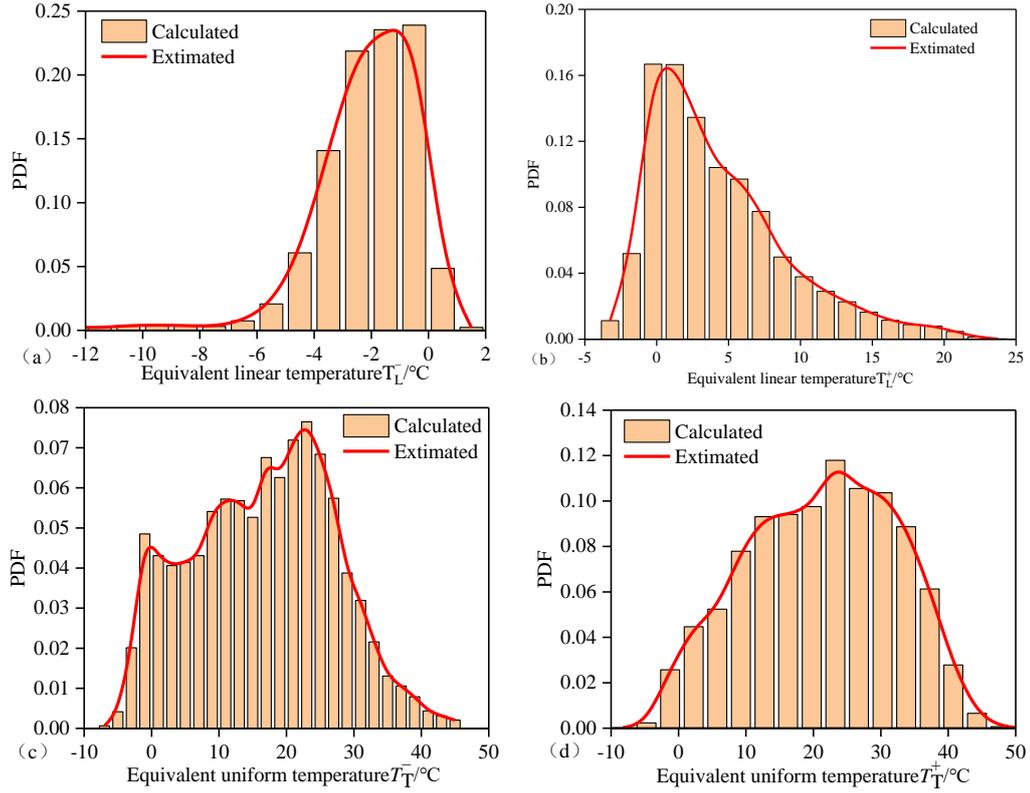

Fig.13. Probability density histogram and estimated curve: (a) negative equivalent linear temperature, (b) positive equivalent linear temperature, (c) negative equivalent uniform temperature, and (d) positive equivalent uniform temperature.

Table 3. Parameters of GMM

| Type | $a_1$ | $b_1$ | $\mu_1$ | $A_2$ | $b_2$ | $\mu_2$ | $A_3$ | $b_3$ | $\mu_3$ | $a_4$ | $b_4$ | $\mu_4$ | $a_5$ | $b_5$ | $\mu_5$ |
|---|---|---|---|---|---|---|---|---|---|---|---|---|---|---|---|
| $T_T^+$ | 1.54 | 8.88 | 18.03 | −1.47 | 8.53 | 17.91 | −0.02 | 5.17 | 27.03 | 0.04 | 3.61 | 0.11 | 0.004 | 7.09 | 36.66 |
| $T_L^+$ | 0.14 | 0.75 | −0.67 | 0.22 | 2.01 | −2.20 | 0 | 0 | 0 | 0 | 0 | 0 | 0 | 0 | 0 |
| $T_T^-$ | −0.41 | 17.89 | 17.32 | 0.12 | 10.88 | 30.92 | 0.48 | 16.45 | 16.09 | 0 | 0 | 0 | 0 | 0 | 0 |
| $T_L^-$ | 0.13 | 1.97 | 0.50 | 0.10 | 5.76 | 4.69 | 0 | 0 | 0 | 0 | 0 | 0 | 0 | 0 | 0 |

### 3.3.2 Vertical thermal load model II

The equivalent nonlinear temperature of the composite girder under the most unfavorable case can be calculated based on Section1.1.

Figs. 14–15 present the VTLM II of the composite girders. The profile 1 of vertical thermal load model I and positive thermal load model shown in Fig. 14 (a) is equivalent to the thermal effects of the composite girder and the same as the profile 2 of vertical thermal load model I and positive thermal load model (Fig. 14 (b)). The values of equivalent linear temperature decomposed in the PTLM of VTLM II and the NTLM of VTLM II are 27.5°C and −12.3°C, respectively. The induced thermal linear slope *k* values are 18.9 and −12.3°C/m, respectively, which can be obtained from the process in Section 3.3.1. The NTLM of VTLM II leads to negative linear temperature distribution, negative induced thermal linear slope, and downward deformation in bridges. Meanwhile, the results of PTLM of VTLM II are opposite. This finding is consistent with the effects of VTLM I.

Fig. 15 displays the simplified schematic of VTLM II, and the parameters of VTLM II are given in Table 4 to facilitate its use in the design of composite girder bridges. The PTLM and NTLM of VTLM II consist of $T_T$, $T_L$, and $T_{NL}$ (Figs. 8 (a–b)). VTLM II is more conducive to explaining the influence of thermal load on the bridge structure from the perspective of mechanism. Fig. 16 shows the relationship between the thermal effect and thermal loads based on VTLM I and VTLM II toward the steel bridge and concrete bridge in service. VTLM II is more convenient to calculating the thermal effect from the secondary effect and the spontaneous effect than VTLM I.

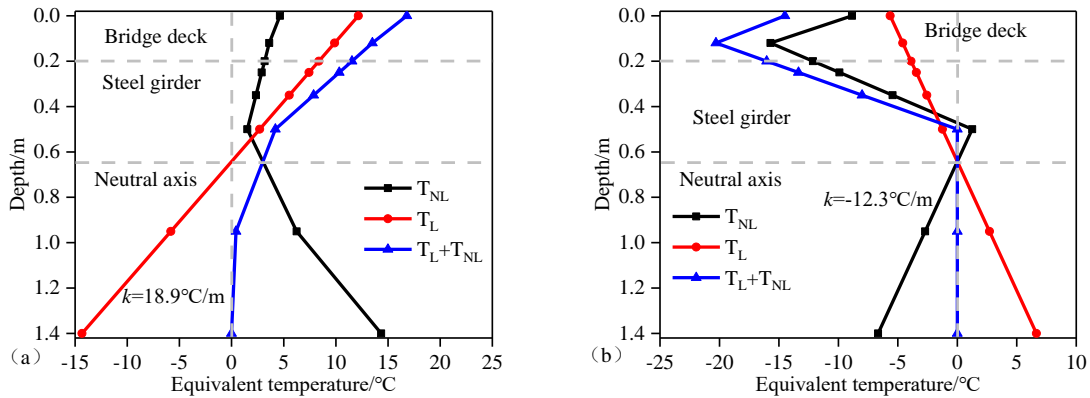

Fig. 14. Vertical thermal load model II: (a) positive thermal load model and (b) negative thermal load model.

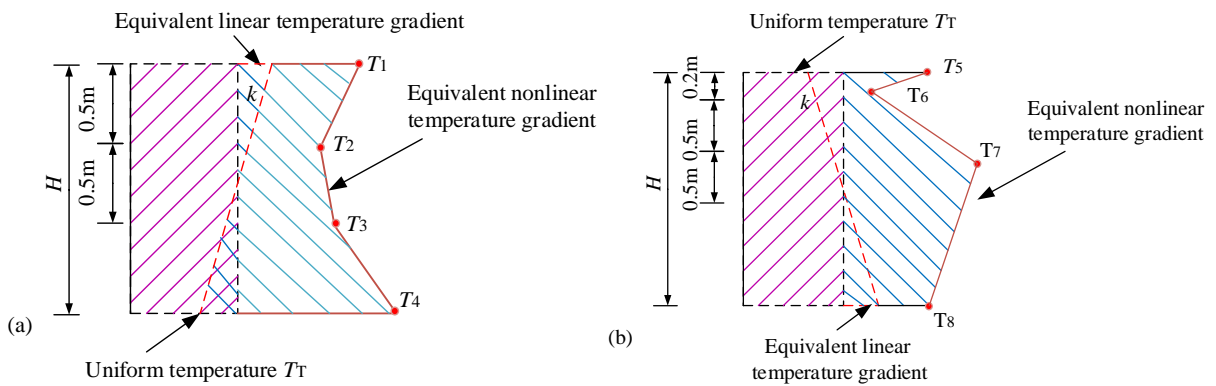

Fig. 15. Simplified schematic of vertical thermal load model II: (a) positive thermal load model and (b) negative thermal load model.

Table. 4 Parameters in vertical thermal load model II.

| Positive thermal load model | | | | | | Negative thermal load model | | | | | |
|---|---|---|---|---|---|---|---|---|---|---|---|
| $T_1$/°C | $T_2$/°C | $T_3$/°C | $T_4$/°C | $k$/°C/m | $T_T$/°C | $T_5$/°C | $T_6$/°C | $T_7$/°C | $T_8$/°C | $k$/°C/m | $T_T$/°C |
| 4.6 | 1.5 | 6.2 | 14.3 | 18.9 | 52.4 | −8.9 | −13.8 | 1.3 | −6.7 | −12.3 | −13.5 |

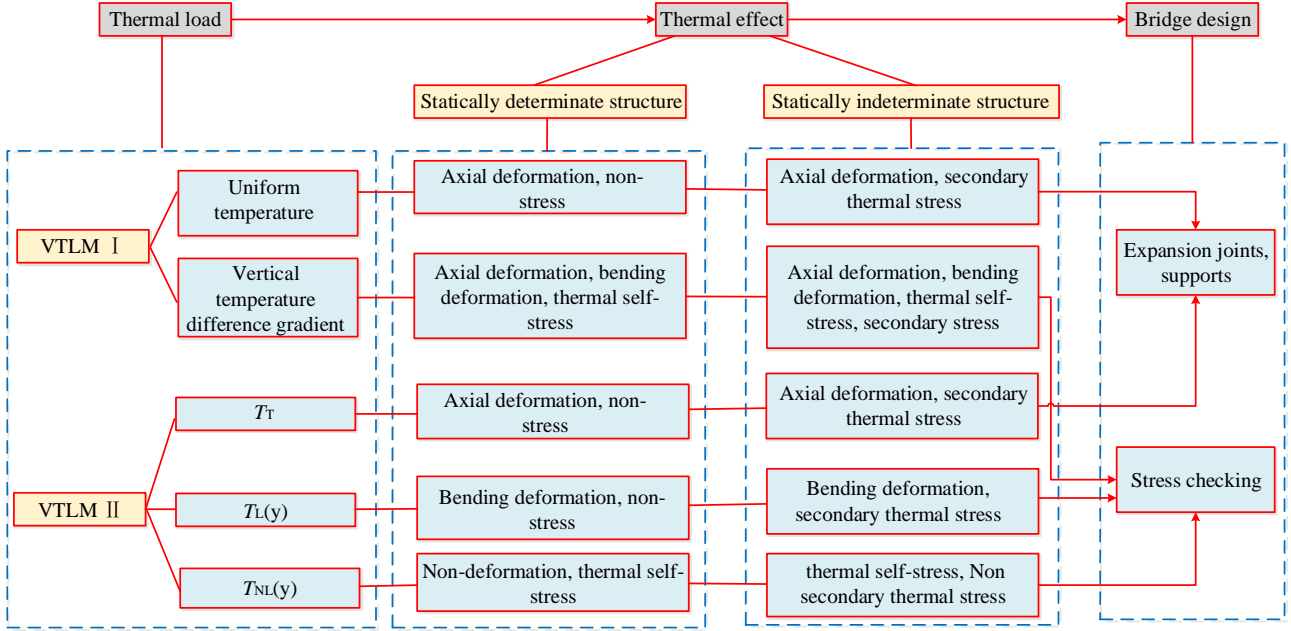

Fig. 16 Relationship between the thermal effects and thermal loads.

## 4 Thermal stress of the composite girder

### 4.1 Calculation method of thermal stress

Taking the centroid of the converted section of the composite girder as the coordinate origin, the equation can be obtained:

$$\int_{h_1}^{h_2} E(y) y b(y) dy = 0 \tag{29}$$

Considering the fiber constraint and the structural boundary constraint, we can express the secondary axial force and the secondary bending moment of the steel-concrete composite girder as $N_{con}$ and $M_{con}$, respectively. After substituting $N=N_{con}$ and $M=M_{con}$ into equations (6) and (7), $\varepsilon_0$ and $\varphi_0$ can be obtained as follows:

$$\varepsilon_0 = \frac{N_{con} + N_0}{EA} \tag{30}$$

$$\varphi_0 = \frac{M_{con} + M_0}{EI} \tag{31}$$

where $N_0 = E_c \alpha_c T_1 + E_s \alpha_s T_2$, $M_0 = E_c \alpha_c T_3 + E_s \alpha_s T_4$, $EA = E_c A_c + E_s A_s$, $EI = E_c I_c + E_s I_s$, $T_1 = \int_{A_c} T(y) dA$,

$T_2 = \int_{A_s} T(y)dA$, $T_3 = \int_{A_c} T(y)ydA$, $T_4 = \int_{A_s} T(y)ydA$, $I_c$, and $I_s$ are the inertia moment of the concrete and steel girders related to the converted section axis.

Generally, conventional bridges obey the axial free expansion and deformation of girders under the thermal effect; hence, $N_{con} = 0$. After substituting equations (30) and (31) into equation (5), the overall thermal stress at any position can be obtained by equation (32):

$$\sigma_{sum}(y) = \frac{E(y)N_0}{EA} + \frac{E(y)y(M_0 + M_{con})}{EI} - E(y)\alpha(y)T(y). \tag{32}$$

When $M_{con} = 0$ is set in equation (32), the thermal self-restraint stress can be denoted as follows:

$$\sigma_s(y) = \frac{E(y)N_0}{EA} + \frac{E(y)yM_0}{EI} - E(y)\alpha(y)T(y). \tag{33}$$

Thermal self-restraint stress caused by the equivalent nonlinear temperature gradient can be calculated by equations (20) and (33).

After substituting $T(y) = G(y)$, $T(y) = T_T$, and $T(y) = T_{NL}(y)$ into equation (32), the thermal stress caused by structural boundary constraints can be calculated by equation (34):

$$\sigma_{con} = \frac{E(y)M_{con}y}{EI}. \tag{34}$$

Suppose an ($n$ + 1)-span continuous composite girder, the bending moment caused by the redundant constraints are $P_i (i = 1, 2, ..., n)$, and the bending moment caused by the redundant constraints of ($n$ + 1)-span continuous girder is as follows:

$$M_{con} = \sum_{i=1}^{n} P_i \overline{M_i}. \tag{35}$$

The relative rotation angle $\Delta_P$ at any support $P$ in the direction of redundant force can be expressed as equation (36) based on the principle of virtual work:

$$\Delta_P = \int_L \overline{M_P} \varphi_0 dx, \tag{36}$$

where $\overline{M_P}$ is the reaction moment induced by unit rotation at support $P$.

Equations (31) and (34) are substituted into equation (35). $\Delta_P = 0$ can be obtained based on boundary coordination, so equation (35) can be calculated as follows:

$$\Delta_P = \sum_i^n \int_L \frac{\overline{M_i}\overline{M_P}}{EI} P_i dx + \int_L \frac{M_0 \overline{M_P}}{EI} dx = 0, \quad (i = 1, 2, ..., n). \tag{37}$$

The secondary bending moment can be obtained by solving equation (37).

**4.2 Thermal stress analysis**

The span arrangement of the example bridge is (2 × 30) m, and the sectional dimensions of the continuous steel-concrete composite girder are shown in Fig. 1. The profile 1 of VTLM I and PTLM of VTLM II are vertical thermal load model for positive temperature effects, and the profile 2 of VTLM I and NTLM of VTLM II are profiles for negative temperature effects. The thermal stress under the thermal effect of VTLM and vertical temperature difference gradient in CS are also calculated. Fig. 17 displays the vertical distribution of thermal self-stress and secondary thermal stress in the mid-support section of the steel-concrete composite girder. The PTLM of VTLM II and NTLM of VTLM II calculates the stress from equivalent linear temperature gradient and equivalent nonlinear temperature gradient without considering the equivalent uniform temperature. As shown in Figs. 17 (a–b), the differences between thermal self-stress and secondary thermal stress under the effect of VTLM I and VTLM II are small, indicating that the thermal effects generated by VTLM I and VTLM II are consistent. For the profile 2 of VTLM I and the NTLM of VTLM II, the thermal self-stress at the top surface of the concrete bridge deck is 1.478 and 1.331 MPa, respectively, which are larger than the 1.183 MPa recommended in CS. For the profile 1 of VTLM I and the PTLM of VTLM II, thermal self-stress at the top of the bridge deck is −0.604 and −0.812 MPa, respectively, which are less than the −1.452 MPa recommended in CS. Large thermal self-stress is found in the position of the inflection part of VTLM, and the thermal self-stress caused by VTLM I and VTLM II is larger than that of the recommended value in CS for steel girder. The secondary thermal stress presents linear distribution (Fig. 9 (b)), and the secondary thermal stresses of the concrete deck and steel girder are larger than the recommended value in CS in VTLM I and VTLM II. If biased vertical temperature difference distribution in CS is applied for the composite girder, then the thermal secondary stress will deviate, which is unfavorable for the accurate design of the composite girder. Therefore, safety can be appropriately enhanced when thermal stress is calculated according to a single standard in the design stage.

According to the above analysis, the thermal stress caused by the vertical temperature difference gradient in CS can hardly surpass the thermal stress caused by temperature loads in the design service life of the composite girder. Particularly, large deviations exist between VTLM and the vertical temperature difference gradient in CS, leading to the thermal deformations of the girder or the thermal stresses of the steel girder being greater than the design values. The VTLM I and VTLM II can replenished the thermal loads of composite girders in the design stage. Additionally, the proposed thermal load decomposition method and thermal stress calculation method can enrich the calculation theory of bridge temperature effects. This work provides a calculation method for the static effect of the bridge thermal loads and a calculation idea for the calculation of the bridge fatigue damage under temperature effects.

Figs. 18 (a–b) show the time history of thermal self-stress of the bridge deck and steel girder roof (the position of the maximum stress value) under the effect of equivalent uniform temperature. The maximum and minimum thermal self-stresses on the top surface of the bridge deck are 0.714 and −0.119 MPa, respectively, which exceed the 0.798 and −0.206 MPa calculated by the $T_T$ (PTLM) of VTLM II and the $T_T$ (NTLM) of VTLM II. The maximum/minimum thermal self-stresses on the bottom surface of the bridge deck are 1.124 and −0.187 MPa, respectively, which are smaller than the 1.256 and −0.323 MPa obtained in the $T_T$ (PTLM) of VTLM II and the $T_T$ (NTLM) of VTLM II. The maximum and minimum self-stresses of the steel girder are 2.099 and −12.623 MPa, respectively, which are less than the 3.631 and −14.094 MPa calculated by the $T_T$ (PTLM) of VTLM II and the $T_T$ (NTLM) of VTLM II, respectively. Hence, the $T_T$ (PTLM) of VTLM II and the $T_T$ (NTLM) of VTLM II can be considered a good method to calculate the thermal stress produced by uniform temperature.

As shown in Figs. 19 (a–b), secondary thermal stress exhibits a cycle characteristic due to the daily and annual cycle of temperature. The secondary thermal stress is usually larger than the thermal self-stress (Figs. 19 (c–d)) in the composite girder. The maximum secondary thermal stress caused by $T_T$, $T_L$, and $T_{NL}$ for the top surface of the bridge deck is 0.328, 6.373, and 0.141 MPa, respectively, and the corresponding minimum value is −1.973, −5.177, and −0.126 MPa, respectively. The maximum secondary thermal stress caused by $T_T$, $T_L$, and $T_{NL}$ for the top surface of the steel girder is 13.909, 36.465, and 0.884 MPa, respectively, and the corresponding minimum value is −2.31, −44.899, and −0.993 MPa, respectively. The equivalent linear temperature $T_L$ is the main factor leading to the secondary thermal stress. Compared with that of $T_L$, the influence of $T_{NL}$ on the secondary thermal stress can be ignored. Equivalent uniform temperature $T_T$ can also lead to the secondary thermal stress for the statically indeterminate structure.

The thermal self-stresses caused by $T_T$, $T_L$, and $T_{NL}$ on the top surface of the bridge deck and steel girder are shown in Figs. 18 and 19 (c–d). The maximum thermal self-stress caused by $T_T$, $T_L$, and $T_{NL}$ for the top surface of the bridge deck is 0.715, 0.251, and 3.759 MPa, respectively, and the corresponding minimum value is −0.118, −0.205, and −2.794 MPa, respectively. For steel girder, the maximum thermal self-stress caused by $T_T$, $T_L$, and $T_{NL}$ is 2.099, 2.100, and 13.040 MPa, respectively, and the corresponding minimum value is −12.623, −2.586, and −32.589 MPa, respectively. Therefore, the equivalent nonlinear temperature $T_{NL}$ is the main factor causing the thermal self-stress. Compared with that of $T_{NL}$, the influence of equivalent linear temperature $T_L$ and equivalent uniform temperature $T_T$ on thermal self-stress can be ignored.

Based on the above discussion, VTLM II composed of $T_T$, $T_L$, and $T_{NL}$ is conducive to calculating the thermal stress of the composite girders from two aspects (secondary thermal stress and spontaneous thermal stress). First, the contribution of different temperature components to the secondary stress and spontaneous stress is easy to obtain. Second, the relationship between the thermal effects and thermal loads is suitable for steel-concrete composite girders (Fig. 16). The proposed thermal load model is expected to have guiding significance for improving the design safety of steel-concrete composite bridges.

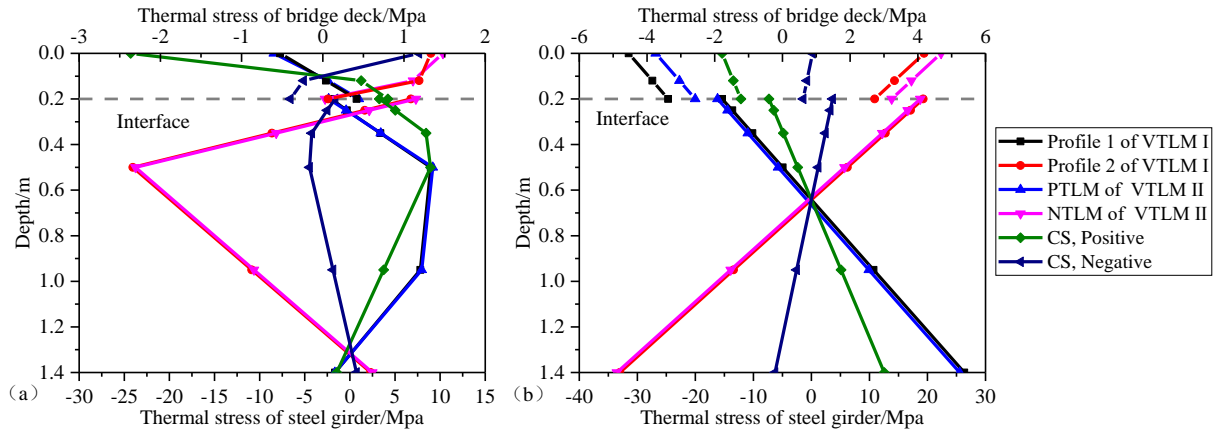

Fig. 17. Thermal stress of the composite girder: (a) thermal self-stress and (b) secondary thermal stress.

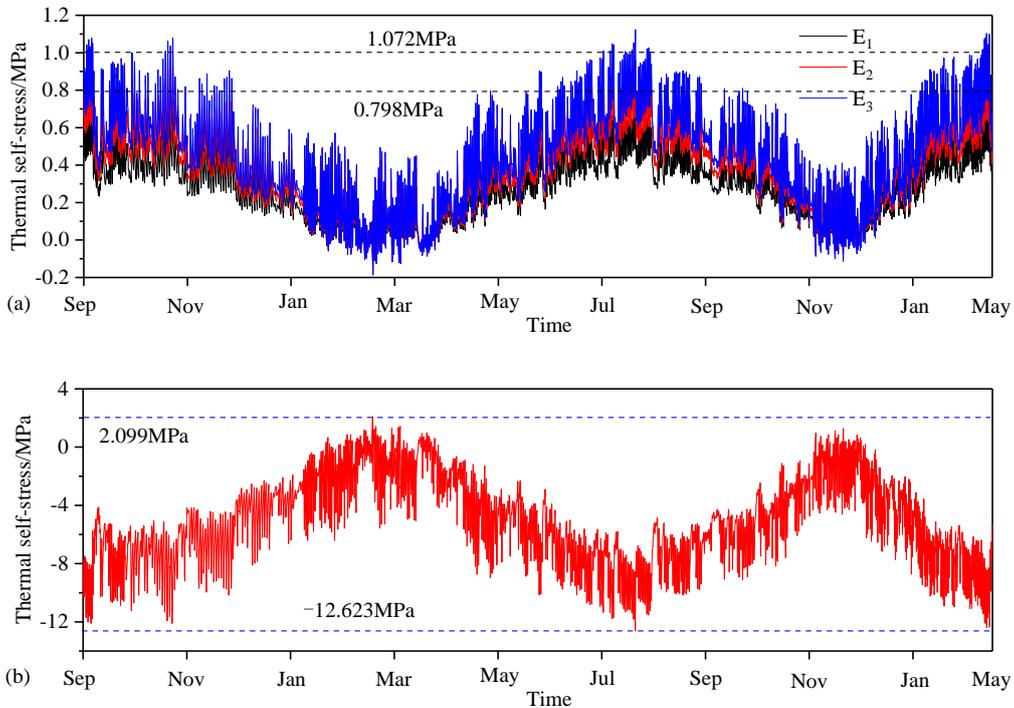

Fig. 18. Time history of thermal self-stress: (a) stress of the bridge deck and (b) stress of the steel girder roof.

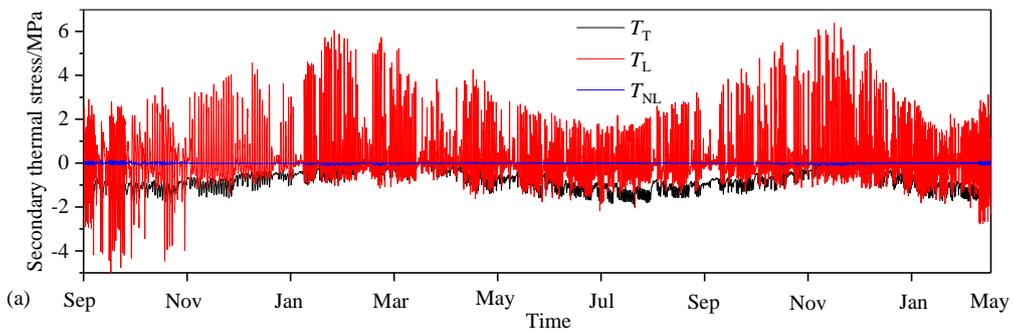

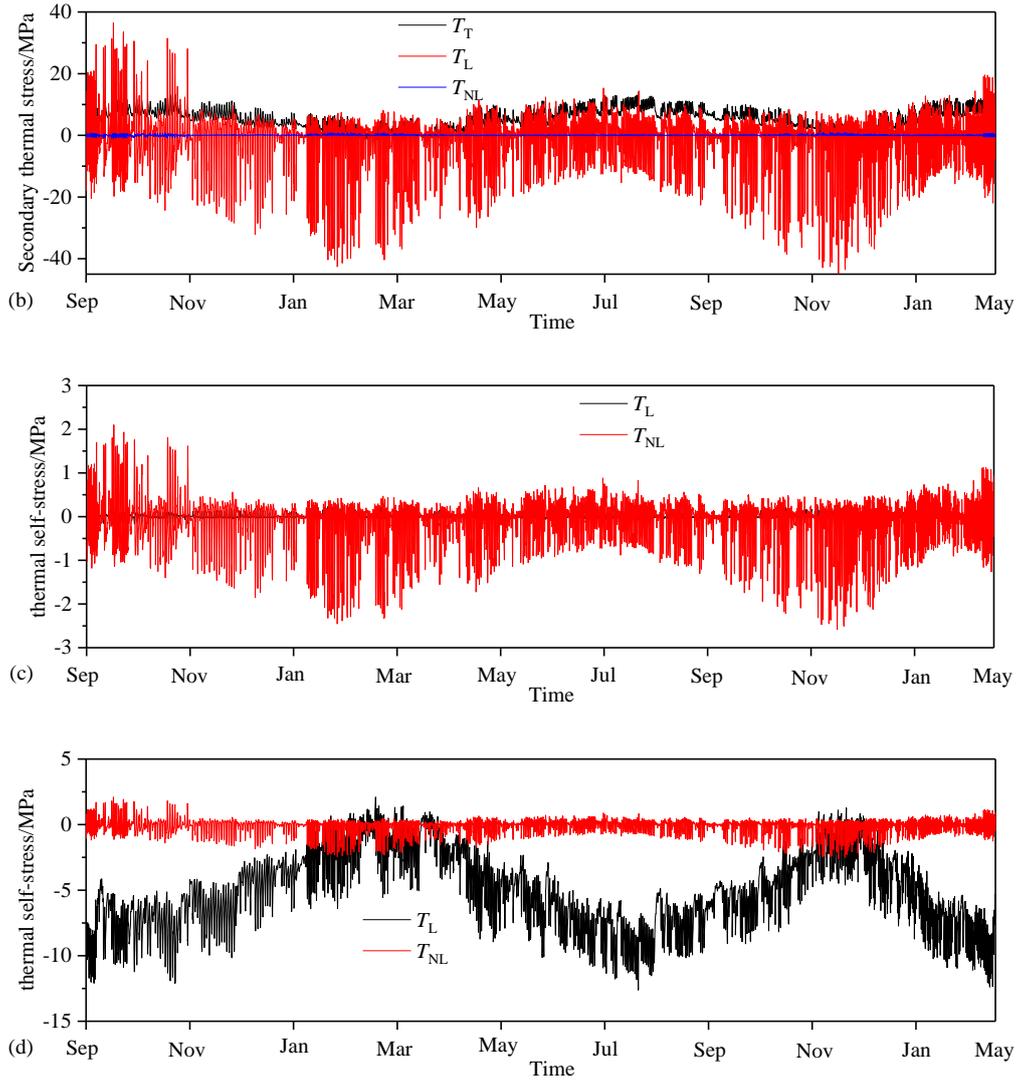

Fig. 19. Time history of thermal stress: (a) self-stress of the bridge deck, (b) self-stress of the steel girder roof, (c) secondary stress of the bridge deck, and (d) secondary stress of the steel girder roof.

## 5 Conclusions

To consider temperature effects and mechanical mechanism in designing the thermal load model of the composite girders, this work proposed a calculation method for decomposing vertical nonlinear temperature. Stand temperature difference and stand equivalent temperature ($T_T$, $T_L$, and $T_{NL}$) were analyzed, and two thermal load models were obtained by calculating and analyzing the temperature field information of the composite girder. The following conclusions can be drawn as follow:

(1) The calculating method of decomposing vertical nonlinear temperature was set up based on the thermal effects equivalence method. Furthermore, the calculation method of thermal stress in composite girders was proposed. The vertical nonlinear temperature was divided into $T_T$, $T_L$, and $T_{NL}$. These three parameters were solved by calculating long-term temperature data. Stand values of $T_T$, $T_L$, $T_{NL}$, and temperature difference with a 50-year return period were also obtained by probability analyses.

(2) VTLM I and VTLM II were established carefully. The profile 1 of VTLM I and PTLM of VTLM II are positive temperature gradients, and the profile 1 of VTLM I can be described with a three-segment linear function composed of $T_{P1}$, $T_{P2}$, and $T_{P3}$. The PTLM of VTLM II has $T_T$, $T_L$, and $T_{NL}$ with a positive temperature effect. Additionally, the profile 2 of VTLM I and NTLM of VTLM II are negative temperature gradients, and the profile 2 of VTLM I can be expressed by a two-segment linear function with parameters $T_{N1}$ and $T_{N2}$. The NTLM of VTLM II is composed of $T_T$, $T_L$, and $T_{NL}$ with a negative temperature effect. VTLM II can clearly reflect the contribution of $T_T$, $T_L$, and $T_{NL}$ to the secondary thermal stress and thermal self-stress of the composite girders.

(3) Extreme values of daily temperature differences $T_{P1}$, $T_{P2}$, $T_{P3}$, $T_{N1}$, and $T_{N2}$ can be described by the normal distribution function, and the stand values of $T_{P1}$, $T_{P2}$, $T_{P3}$, $T_{N1}$, and $T_{N2}$ are 16.8°C, 13.5°C, 4.2°C, −14.5°C, and −20.3°C, respectively, with a 50-year return period. Daily maximum and minimum values of $T_T$ and $T_L$ can be expressed by GMM. The maximum stand values of $T_T$ and $T_L$ are 52.4°C and 27.5°C, respectively, and the minimum are −13.5°C and −12.3°C, respectively, with a 50-year return period.

(4) For the VTLM II of the composite girders, $T_L$ is one of the main factors causing secondary thermal stress and can also cause secondary thermal stress for the statically indeterminate structure. However, the stress caused by $T_L$ and $T_T$ is quite small for the statically determinate structure. $T_T$ is the main factor causing the axial deformation of the composite girder. $T_{NL}$ mainly causes the thermal self-stress of the composite girders, but its contribution to the secondary thermal stress can be ignored. VTLM II can clearly express the relationship between thermal loads and thermal effects.

(5) The cyclic equivalent uniform temperature $T_{CUT}$ is 8376°C. Large deviations exist among VTLM I, VTLM II, and the vertical temperature difference gradients in CS, which may lead to thermal stresses and deformations of composite girders being greater than the calculated values in CS.

# 6 Declaration of competing interest

The authors declare that they have no known competing financial interests or personal relationships that could have appeared to influence the work reported in this paper.

# 7 Acknowledgements

The authors gratefully acknowledge the financial support provided by the National Natural Science Foundation of China (Grant No.51708269) and the Technology Projects of Gansu Province (Grant No. 19ZD2GA002). The constructive comments from the editor and reviewers are also greatly appreciated.## 8 References

[1] Fan J, Li B, Liu C, Liu Y. An efficient model for simulation of temperature field of steel-concrete composite beam bridges. Structures 2022; 43: 1868–1880.

[2] He Z, Ma Z, Zhang S, Liu Z, Ma Z. Temperature gradients and stress distributions in concrete box-girder bridges during hot-mix asphalt paving. Structures 2021; 33: 1954–1966.

[3] Zhang C, Liu Y, Ye F, Liu J, Yuan Z, Zhang G, Ma Z. Validation of long-term temperature simulations in a steel-concrete composite girder. Structures 2020; 27: 1962–1976.

[4] Lawson L, Ryan KL, Buckle IG. Bridge temperature profiles revisited: thermal analyses based on recent meteorological data from Nevada. J Bridge Eng 2020; 25(1): 04019124.

[5] Han Q, Ma Q, Xue J, Xu J, Liu M. Structural health monitoring research under varying temperature condition: a review. J Civ Struct Health 2021; 11: 149–173.

[6] Sousa Tomé E, Pimentel M, Figueiras J. Structural response of a concrete cablestayed bridge under thermal loads. Eng Struct 2018; 176: 652–672.

[7] Gottsäter E, Larsson Ivanov O, Molnár M, Plos M. Validation of temperature simulations in a portal frame bridge. Structures 2018; 15: 341–348.

[8] Nasr A, Kjellström E, Björnsson I, Honfi D, Ivanov OL, Johansson J. Bridges in a changing climate: a study of the potential impacts of climate change on bridges and their possible adaptations. Struct Infrastruct Eng 2020; 16: 738–749.

[9] Silveira AP, Branco FA, Castanheta M. Statistical analysis of thermal actions for concrete bridge design. Struct Eng Int 2000; 10: 33–38.

[10] Westgate R, Koo K-Y, Brownjohn J. Effect of solar radiation on suspension bridge performance. J Bridge Eng 2015; 20(5): 04014077.

[11] Soukhov D. Two methods for determination of linear temperature differences in concrete bridges with the help of statistical analysis. Darmstadt Concr 1994; 9: 193–210.

[12] Hottel HC. A simple model for estimating the transmittance of direct solar radiation through clear atmospheres. Sol Energy 1976; 18: 129–134.

[13] Mirambell E, Aguado A. Temperature and stress distributions in concrete box girder bridges. J Struct Eng
25

1990; 116(9): 2388–409.

[14] Roberts-Wollman CL, Breen JE, Cawrse J. Measurements of thermal gradients and their effects on segmental concrete bridge. J Bridge Eng 2002; 7(3): 166–174.

[15] Zuk W, Simplified design check of thermal stresses in composite highway bridges, Highw. Res. Rec. 1965; (13): 10–13.

[16] Wang C, Duan L, Zhai M, Zhang Y, Wang C, Steel bridge long-term performance research technology framework and research progress, Adv. Struct. Eng. 2016; 20(1): 51–68.

[17] Gottsäter E, Larsson Ivanov O, Molnár M, Plos M. Validation of temperature simulations in a portal frame bridge. Structures 2018; 15: 341–348.

[18] MOT (Ministry of Transport of the People's Republic of China). General specifications for design of highway bridges and culverts. JTG D60-2015. Beijing, China; 2015.

[19] AASHTO (American Association of State Highway and Transportation Officials). AASHTO LRFD bridge design specifications. 8th ed. Washington, DC; 2017.

[20] Maguire M, Moen CD, Roberts-Wollmann C, Cousins T. Field verification of simplified analysis procedures for segmental concrete bridges. J Struct Eng 2015; 141(1): D4014007.

[21] Mirambell E, Aguado A. Temperature and stress distributions in concrete box girder bridges. J Struct Eng 1990; 116(9): 2388–2409.

[22] Zheng D, Qian Z, Liu D, Zhang X, Liu Y. Thermal field characteristics of reinforced concrete box girder during high-temperature asphalt pavement paving. Transport Res Rec 2018; 2672(41): 56–64.

[23] Saetta A, Scotta R, Vitaliani R. Stress analysis of concrete structures subjected to variable thermal loads. J Struct Eng 1995; 121(3): 446–457.

[24] Lin J, Briseghella B, Xue J, Tabatabai H, Huang F, Chen B. Temperature monitoring and response of deck-extension side-by-side box-girder bridges. J Perform Constr Facil 2020; 34(2): 04019122.

[25] Larsson O, Thelandersson S. Estimating extreme values of thermal gradients in concrete structures. Mater Struct 2011; 44(8): 1491–500.

[26] Li D, Maes MA, Dilger WH. Thermal design criteria for deep prestressed concrete girders based on data from confederation bridge. Can J Civ Eng 2004; 31(5): 813–825.

[27] Saetta A, Scotta R, Vitaliani R. Stress analysis of concrete structures subjected to variable thermal loads. J Struct Eng 1995; 121(3): 446–457.

[28] Maher D. The effects of differential temperature on continuous prestressed concrete bridges. Inst Engrs Civil

Eng Trans 1970; 12(1): 29–32.

[29] Shen R, Du M, Jiang Y. Study on temperature load pattern in double-box single-cell steel box girder during asphalt concrete paving at high temperature. Rail Eng 2019; 59(4): 38–43.

[30] Lei X, Fan X, Jiang H, Zhu K, Zhan H. lateral temperature gradient effect on a PC box-girder bridge based on real-time solar radiation and spatial temperature monitoring. Sensors 2020; 20(18): 5261.

[31] Cai C, Huang S, He X, Zhou T, Zou Y. Investigation of concrete box girder positive temperature gradient patterns considering different climatic regions. Structures 2022; 35: 591–607.

[32] Abid SR, Cacciola P. Three-dimensional finite element temperature gradient analysis in concrete bridge girders subjected to environmental thermal loads. Cogent Eng 2018; 5(1): 1447223.

[33] Tian Y, Zhang N, Xia H. Temperature effect on service performance of high-speed railway concrete bridges. Adv Struct Eng 2017; 20(6): 865–883.

[34] Abid SR. Temperature variation in steel beams subjected to thermal loads. Steel Compos Struct 2020; 34: 819–835.

[35] Mussa FI, Abid SR. Investigation of temperature gradients in composite girders in the southern region of the black sea. J Phys Conf Ser 2021; 1895(1): 012069.

[36] Wang G, Ding Y. Reliability estimation of horizontal rotation at beam end of long-span continuous truss bridge affected by temperature gradients. J Perform Constr Fac 2019; 33(6): 04019061.

[37] Liu J, Liu YJ, Bai YX, Liu GL. Study on regional difference and zoning of the temperature gradient pattern of concrete box girder. China J Highway Transp 2020; 33(3): 25–39.

[38] Fan J, Liu Y, Liu C. Experiment study and refined modeling of temperature field of steel-concrete composite beam bridges. Eng Struct 2021; 240: 112350.

[39] Liu J, Liu Y, Zhang G. Experimental analysis of temperature gradient patterns of concrete-filled steel tubular members. J Bridge Eng 2019; 24(11): 04019109.